\documentclass[rnote]{aa} 

\usepackage[varg]{txfonts}
\usepackage{graphicx}
\usepackage{lscape}
\begin{document}

   \title{The PMS star V1184 Tau (CB 34V) at the end of prolonged eclipse}

     \author{E. H. Semkov
          \inst{1}
          \and
          S. P. Peneva\inst{1}
          \and
          S. I. Ibryamov\inst{1,2}
          }

   \institute{Institute of Astronomy and National Astronomical Observatory,
Bulgarian Academy of Sciences, 72 Tsarigradsko Shose Blvd.,
1784 Sofia, Bulgaria\\
              \email{esemkov@astro.bas.bg}
         \and
Department of Theoretical and Applied Physics, Faculty of Natural Sciences, University of Shumen, 115, Universitetska Str., 9712 Shumen, Bulgaria}

   \date{Received  ; accepted  }

 
  \abstract
   {}
   {V1184 Tau (CB 34V) lies in the field of the Bok globule CB 34 and was discovered as a large amplitude variable in 1993.
According to the first hypothesis of the variability of the star, it is a FU Orionis candidate erupted between 1951 and 1993.
During subsequent observations, the star manifests large amplitude variability interpreted as obscuration from circumstellar clouds of dust.
We included V1184 Tau (CB 34V) in our target list of highly variable pre-main-sequence stars  to determine the reasons for the variations in the brightness of this object.}
   {Data from $BVRI$ photometric observations of the young stellar object V1184 Tau, obtained in the period 2008$-$2015, are presented in the paper.
These data are a continuation of our optical photometric monitoring of the star began in 2000 and continuing to date.
The photometric observations of V1184 Tau were performed in two observatories with two medium-sized and two small telescopes.}
   {Our results indicate that during periods of maximum light the star shows characteristics typical of T Tauri stars. During the observed deep minimum in brightness, however, V1184 Tau is rather similar to UX Orionis objects.
The deep drop in brightness began in 2003 ended in 2015 as the star has returned to maximum light.
The light curve during the drop is obviously asymmetric as the decrease in brightness lasts two times longer than the rise.
The observed colour reverse on the colour-magnitude diagrams is also confirmation of obscuration from circumstellar clouds of dust as a reason for the large amplitude variability in the brightness.}
   {}


   \maketitle
%

\section{Introduction}

Photometric variability is a fundamental characteristic of all types of pre-main-sequence (PMS) stars.
In many cases, the discovery of variability is a proof for the membership of stars in a certain group of young objects. 
Depending on the mass, PMS stars are divided in two main groups: the widely distributed T Tauri stars (TTS), determined as young low-mass stars ($\it M$ $\leq$ $2M_{\sun}$) with emission line spectra and irregular photometric variability; and the more massive ($M$ $>$ $2M_{\sun}$) Herbig Ae/Be stars (HAEBE).
Respectively TTS can be divided in two subgroups: classical T Tauri (CTT) stars surrounded by massive accretion circumstellar disks and weak line T Tauri (WTT) stars without indications of accretion from the disk (Bertout \citeyear{ber}). 

In accordance with \citet{her07} the photometric variability of WTT stars is caused by the presence of large cool spots or groups of spots analogous to sunspots.
The periods of variability on timescales of days and amplitudes up to 0\fm6 ($I$) are typical of the WTT stars variability. The brightness variations of CTT stars are more complicated: the variability is produced by mixture of cool and hot surface spots with non-periodic variations and amplitudes up to 2-3 mag. ($I$).

The very rare phenomena in PMS evolution, i.e. large amplitude, long lasting eruptions, are grouped into two main types: FU Orionis (FUor) and EX Lupi.
The outburst of FUor objects usually continues over several decades, and the time of the rise of brightness is less than the time of the decline (see Audard et al. \citeyear{aud}, Reipurth \& Aspin \citeyear{ra10} and references therein).
Well-studied FUors have the following same characteristics:  a $\Delta$$V$$\approx$4-6 mag. amplitude of the outburst, connection with reflection nebulae, belonging to regions of star formation, an F-G supergiant spectrum in the optical range, a P Cyg profile of H$\alpha$ line and Na I doublet, a strong LiI~6707~\AA\ absorption line, and CO bands in near-infrared spectra (Herbig \citeyear{herb}).  

A sizable number of HAEBE stars with spectral type later than A0 and some early F-G types TTS exhibit very strong photometric variability with abrupt quasi-Algol drops in brightness with amplitudes reaching to 3 mag. ($V$) (Nata et al. \citeyear{nata}).
During the very deep minima in brightness an increasing of polarization and peculiar colour variability are observed.
The prototype of this group of PMS stars named UXors is UX Orionis.
The widely accepted explanation of its variability is a variable extinction from orbiting circumstellar clumps or clouds of dust or from edge-on circumstellar disks (Grinin et al. \citeyear{gr91}).

The unique PMS star V1184 Tau (also known as CB 34V) discovered by \citet{yun} is located in the field of the Bok globule CB 34 (Clemens \& Barvainis \citeyear{clem}).
The comparison of CCD images that Yun et al. \citeyear{yun} obtained in 1993 with plates from Palomar Observatory Sky Survey obtained in 1951 reveals an increase in brightness of this star at 3\fm7 ($R$).
The first supposition of \citet{yun} about the nature of V1184 Tau is that it exhibits a FUor type of outburst in optical wavelengths.
\citet{alves} defined the spectral type of V1184 Tau as G5 (III-IV), the mass of the star $\sim$2 M$_{\sun}$, and the age $\sim$$10^6$ yrs.
\citet{tac} discovered a 2.372 days rotation period of V1184 Tau, suggesting the presence of large cool spots on the stellar surface.

The data presented in this paper are a continuation of the optical photometric monitoring of V1184 Tau began in 2000.
The photometric and spectroscopic study presented in our first paper (Semkov \citeyear{sem03}) reveals V1184 Tau as a possible WTT star with an amplitude of $0\fm5$ ($I$) and spectral variability.
In our second paper (Semkov \citeyear{sem04}), the beginning of a new very deep minimum in brightness of V1184 Tau was reported.
In the third paper from our study (Semkov \citeyear{sem06}), new data from optical photometry and spectroscopy of V1184 Tau in the period during the deep minimum are reported.
In  \citet{sem08}, data from archival photographic plates are presented, which proves that an unknown minimum of brightness exists during the approximate period 1980-1985.
Optical photometric data obtained at the time of our photometric monitoring have been published in \citet{tac}, \citet{bar}, and \citet{gr09}.

\section{Observations}

The photometric observations of V1184 Tau were performed in two observatories with four telescopes: the 2-m Ritchey-Chr\'{e}tien-Coud\'{e}, the 50/70-cm Schmidt and 60-cm Cassegrain telescopes of the Rozhen National Astronomical Observatory (Bulgaria), and the 1.3-m Ritchey-Chr\'{e}tien telescope of the Skinakas Observatory\footnote{Skinakas Observatory is a collaborative project of the University of Crete, the Foundation for Research and Technology, Greece, and the Max-Planck-Institut f{\"u}r Extraterrestrische Physik, Germany.} of the University of Crete (Greece).

The observations were performed with five types of CCD cameras: Vers Array 1300B at the 2-m RCC telescope, ANDOR DZ436-BV at the 1.3-m RC telescope, SBIG STL-11000M and FLI PL16803 at the 50/70-cm Schmidt telescope, and FLI PL9000 at the 60-cm Cassegrain telescope. 
The technical parameters for the CCD cameras used, observational procedure, and data reduction process are described in \cite{ibr}.
All frames were taken through a standard Johnson-Cousins set of filters. 
To minimize the light from the surrounding nebula all data were analysed using the same aperture, which was chosen as 4\arcsec radius.
The background is taken from 15\arcsec to 20\arcsec.
As a reference the $IRVB$ comparison sequence reported in \cite{sem03,sem06} was used. 

\section{Results and discussion}

\begin{figure*}
\centering
\includegraphics{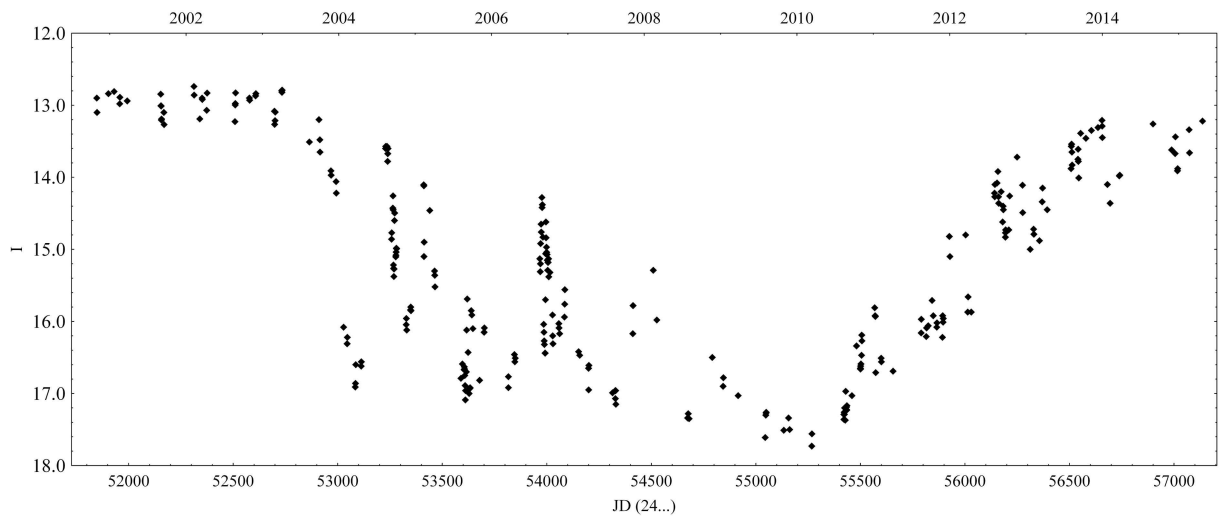}
\caption{$I$ light curve of V1184 Tau for the period October 2000 $-$ April 2015.}
\label{Fig1}%
\end{figure*}

The new data from our optical photometric observations of V1184 Tau are presented in Table~1. 
The columns contains the Date and Julian Date (J.D.) of observation, $IRVB$ magnitudes of V1184 Tau, the telescope and CCD camera used. 
The mean values of instrumental errors of our photometric study are listed in \citet{sem08}.
In very deep minimums the star can be observed only in $V$, $R$, and $I$ bands with the middle size telescopes (2-m RCC and 1.3-m RC) and only in $R$ and $I$ bands with the small telescopes (50/70-cm Schmidt and 60-cm Cassegrain).
The $I$ light curve of V1184 Tau from all our observations is shown in Fig~\ref{Fig1}. 
The error bars are not shown in the figure, as their size is comparable to the symbols used.

The data collected during our long-term photometric monitoring suggest that from October 2000 to April 2003 the brightness of V1184 Tau varies with an amplitude of about 0\fm5 ($I$) without increasing or decreasing.
Since August 2003 a gradual decrease in brightness has begun and the magnitude of the star decreased with $\sim$$4^{\rm m}$ ($I$) until March 2004.
Over the next six years, the brightness of the star  changed rapidly, and several rises and drops with amplitudes of several magnitudes are observed.
The minimum brightness of V1184 Tau ($I$ = $17\fm73$) was registered in March 2010.
Therefore, the process of decrease in brightness continues nearly seven years.

Since the summer of 2010 the brightness of the star began to rise gradually until the spring of 2015 when it reached values close to the maximum light.
Process of increasing in brightness is relatively fast and with a relatively smaller brightness variations, unlike the process of decreasing in brightness.
Therefore, the deep minimum in brightness of V1184 Tau lasts around 12 years and the registered amplitude is $\Delta$$I$$\approx$$4\fm8$.

\begin{figure*}
\centering
\includegraphics{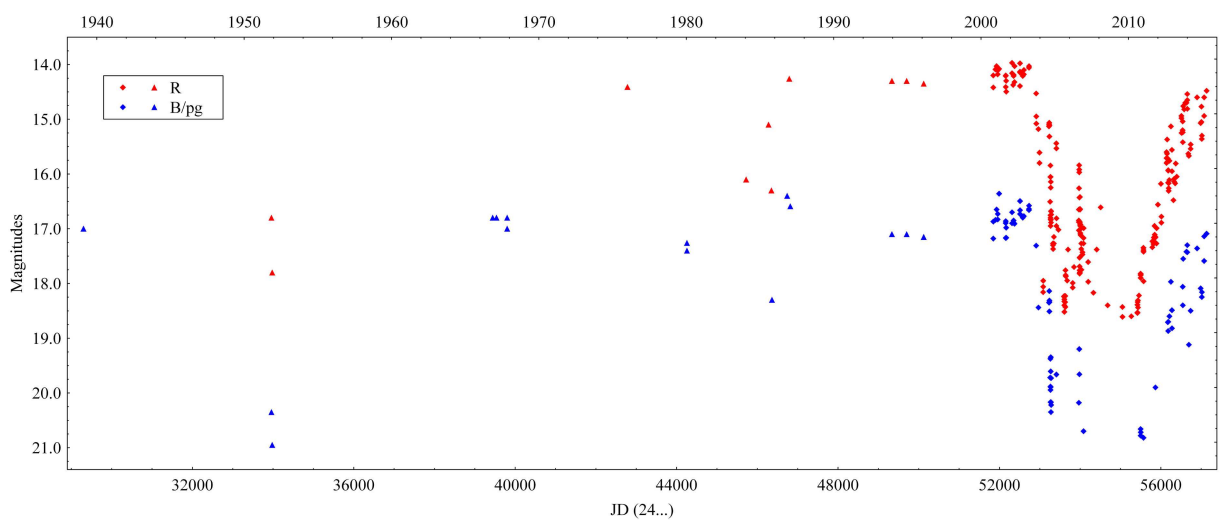}
\caption{Historical $R$ and $B/pg$ light curves of V1184 Tau.}
\label{Fig2}%
\end{figure*}

Fig~\ref{Fig2} shows the long-term $B/pg$ and $R$ light curves of V1184 Tau from all available observations. 
The diamonds denote our CCD photometric data; triangles denote the photographic data collected from archives of several telescopes (Semkov et al.  \citeyear{sem08}).  
The figure shows that a total of three deep minimum in brightness have been registered to date: the first around 1951, the second between 1976 to 1985, and the last  from 2003 to 2015.
Duration of the first minimum documented on the photographic plates from Palomar Observatory Sky Survey is still undetermined, while the duration of the second minimum ranges between six and nine years.
Detailed photometric study of the third minimum indicates a complex asymmetrical light curve, which is  evidence of the irregular structure of the eclipsing object. 

\begin{figure*}
   \centering
   \includegraphics{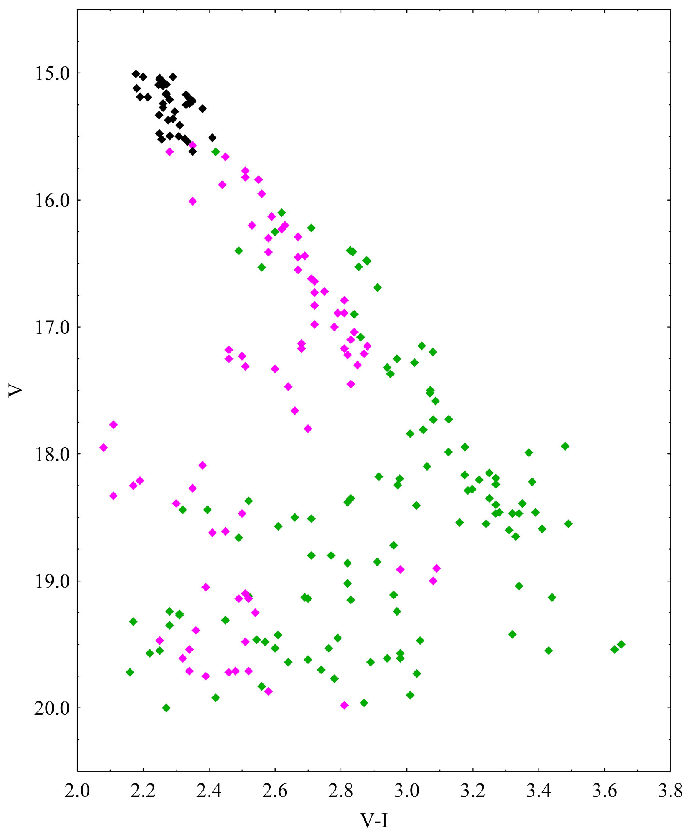}
   \includegraphics{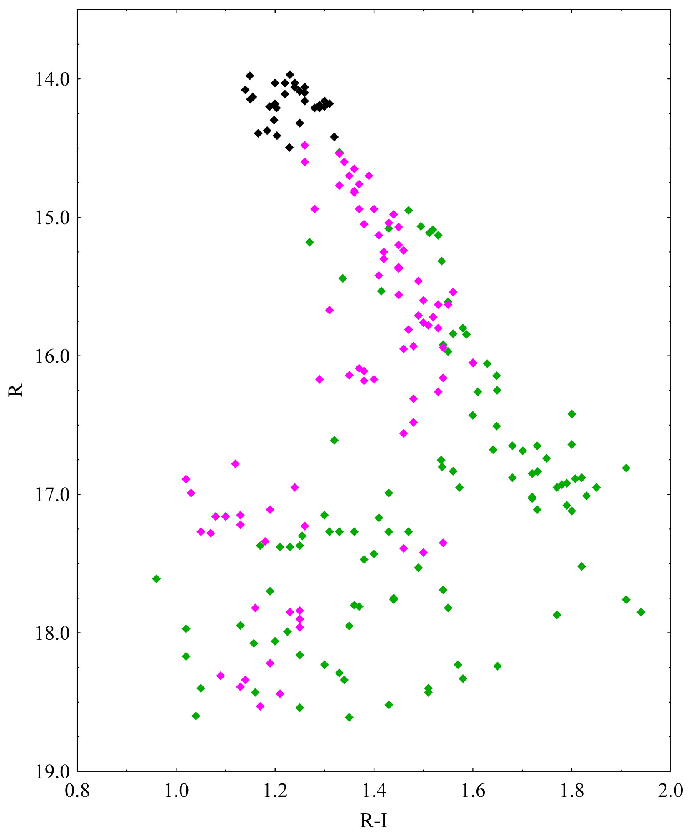}
   \caption{$V/V-I$ and $R/R-I$ diagrams for the period of our photometric observations. In the figure, the black diamonds denote observations from October 2000 to April 2003 (maximum level of brightness out of obscuration), the green diamonds observations from August 2003 to March 2010 (when the brightness of the star dropped) and the pink diamonds from August 2010 to April 2015 (when the brightness of the star increased).} 
   \label{Fig3}
   \end{figure*}

The multi-colour photometry obtained during the deep minimum allows us to examine the variation of colour indices with stellar brightness.
In Fig~\ref{Fig3} we plot the measured colour indices $V-I$ versus $V$ and $R-I$ versus $R$ stellar magnitude.
The results from our study indicate that the star is not obscured by a compact object, but rather from large and dense clouds of dust.
Also, our data indicate that the star has been relatively redder during the decrease in brightness than during the rise in brightness.

During the maximum light the colour variability (star becomes redder as it fades) is typical for WTT stars with large cool spots. 
This variability is caused by the rotation of the spotted surface. 
During the beginning of the eclipse the star also becomes redder, when its light is covered by dust clouds on the line of seeing.
In this case, we assume that the reddening of the star is caused by the variable extinction from the circumstellar environment.
However, from a certain magnitude called turning point (in this case V $\sim$ 18 mag.) the colour of the star becomes bluer.
The reason for this change is that the obscuration rise vastly and the part of the scattered light in the total observed light from the star is significant.

This effect of colour reversal (so-called "blueing") was described in many papers (Bibo \& The \citeyear{bibo}, Grinin et al. \citeyear{gr94}).
In accordance with the model of dust clumps obscuration, the observed colour reversal is produced by the scattered light from the small dust grains.
Our photometric data confirm the existence of a blueing effect in the colour/magnitude diagrams of V1184 Tau, a typical feature of the PMS stars from UXor type.
This is  independent evidence that the variability of V1184 Tau is dominated by the variable extinction from circumstellar environment.
Meanwhile, large cool spots should continue to form on the stellar surface, which additionally modify the brightness and the colour of the star.
Therefore, we observe a combination of both types of variability, which contributes to the large deviation of the points on the colour/magnitude diagrams.
The main cause of the observed large deviation of the points, however, should be the strong fluctuations of the scattered light during our long-term photometric monitoring of the extinction event.

Based on data from infrared $JHK$ photometry, \citet{gr09} propose that  the increase in extinction and the drop in star's brightness can be induced by enhanced accretion from the circumstellar disks. 
The lack of a fall in the K band is explained by a disk model with a puffed up inner rim, whose effective width is increased as a result of enhanced accretion rate.
The presence of another reason for stellar variability, i.e. episodic accretion, may also increase the dispersion of points in the $V/V-I$ and $R/R-I$ diagrams.

The results from our 15 years of photometric monitoring show that V1184 Tau has no analogue among the PMS variables.
  A similar amplitude and duration of the decrease in brightness was only observed in the case of CQ Tau (Grinin et al. \citeyear{gr08}). 
Unlike V1184 Tau, which is WTT star, CQ Tau is a typical HAEBE star with a bigger mass and from an earlier spectral type.
The observed  decrease in brightness during the period 2003-2015  lasted somewhat longer than our prediction (Semkov et al. \citeyear{sem08}), which still does not exclude the presence of the periodicity.
The extremely large amplitude,  rapid variability in brightness and  significant deviation of colour indices can only be explained by the presence of several variability mechanisms, with variable extinction the dominant mechanism.
 
\begin{acknowledgements}
The authors thank the Director of Skinakas Observatory Prof. I. Papamastorakis and Prof. I. Papadakis for the award of telescope time. This research has made use of the NASA's Astrophysics Data System Abstract Service, the SIMBAD database, and the VizieR catalogue access tool, operated at CDS, Strasbourg, France.     
\end{acknowledgements}

\onecolumn
\newpage

\section{ONLINE MATERIAL}

\onllongtab{}{
\begin{longtable}{llllllll}
\caption{Photometric CCD observations of V1184 Tau during the period February 2008 - April 2015}\\
\hline
\hline
\noalign{\smallskip}
Date & J.D. (24...) & I & R & V & B & Telescope & CCD\\
\noalign{\smallskip}
\hline
\endfirsthead
\caption{Continued.}\\ 
\hline
\hline
\noalign{\smallskip}
Date & J.D. (24...) & I & R & V & B & Telescope & CCD\\
\noalign{\smallskip}
\hline
\noalign{\smallskip}
\endhead
\hline
\noalign{\smallskip}
\endfoot
\hline
\noalign{\smallskip}
\endlastfoot
12.02.2008 & 54509.332 & 15.29 & 16.61 & - & - & Scm & ST11000\\ 
29.02.2008 & 54526.397 & 15.98 & -     & - & - & Scm & ST11000\\ 
26.07.2008 & 54673.602 & 17.34 & -     & - & - & 1.3m & ANDOR\\ 
29.07.2008 & 54676.605 & 17.28 & -     & - & - & 1.3m & ANDOR\\ 
02.08.2008 & 54680.595 & 17.35 & 18.40 & 19.57 & - & 1.3m & ANDOR\\ 
20.11.2008 & 54791.491 & 16.50 & -     & - & - & Scm & ST11000\\ 
11.01.2009 & 54843.237 & 16.90 & -     & - & - & Scm & ST11000\\ 
12.01.2009 & 54844.310 & 16.78 & -     & - & - & Scm & ST11000\\ 
24.03.2009 & 54915.303 & 17.03 & -     & - & - & Scm & ST11000\\ 
01.08.2009 & 55044.604 & 17.61 & -     & - & - & 1.3m & ANDOR\\\
04.08.2009 & 55047.604 & 17.30 & -     & 19.55 & - & 1.3m & ANDOR\\ 
05.08.2009 & 55048.607 & 17.27 & 18.43 & 19.83 & - & 1.3m & ANDOR\\ 
06.08.2009 & 55049.589 & 17.26 & 18.61 & - & - & 1.3m & ANDOR\\ 
28.10.2009 & 55133.481 & 17.51 & -     & - & - & Scm & FLI\\ 
20.11.2009 & 55156.495 & 17.34 & -     & - & - & Scm & FLI\\ 
26.11.2009 & 55161.520 & 17.50 & -     & 19.92 & - & 2m & VA\\ 
11.03.2010 & 55267.299 & 17.73 & -     & 20.00 & - & 2m & VA\\ 
12.03.2010 & 55268.269 & 17.56 & 18.60 & 19.72 & - & 2m & VA\\ 
12.08.2010 & 55420.600 & 17.29 & 18.54 & 19.87 & - & 1.3m & ANDOR\\ 
13.08.2010 & 55421.603 & 17.36 & 18.53 & 19.75 & - & 1.3m & ANDOR\\ 
14.08.2010 & 55422.603 & 17.26 & 18.39 & 19.72 & - & 1.3m & ANDOR\\ 
15.08.2010 & 55423.616 & 17.29 & -     & 19.61 & - & 1.3m & ANDOR\\ 
16.08.2010 & 55424.610 & 17.20 & 18.34 & 19.54 & - & 1.3m & ANDOR\\ 
19.08.2010 & 55427.617 & 17.37 & -     & 19.71 & - & 1.3m & ANDOR\\ 
21.08.2010 & 55429.607 & 16.97 & -     & 19.48 & - & 1.3m & ANDOR\\ 
24.08.2010 & 55432.559 & 17.22 & 18.31 & 19.47 & - & 1.3m & ANDOR\\ 
25.08.2010 & 55433.576 & 17.19 & -     & 19.71 & - & 1.3m & ANDOR\\ 
26.08.2010 & 55434.564 & 17.23 & 18.44 & 19.71 & - & 1.3m & ANDOR\\ 
27.08.2010 & 55435.575 & 17.17 & -     & 19.98 & - & 1.3m & ANDOR\\ 
20.09.2010 & 55459.595 & 17.03 & 18.22 & 19.39 & - & 1.3m & ANDOR\\ 
12.10.2010 & 55481.545 & 16.34 & -     & -     & - & 1.3m & ANDOR\\ 
29.10.2010 & 55499.471 & 16.65 & 17.90 & 19.14 & - & 2m & VA\\ 
31.10.2010 & 55500.536 & 16.66 & 17.82 & 19.05 & 20.66 & 2m & VA\\ 
31.10.2010 & 55501.406 & 16.62 & 17.85 & 19.14 & 20.78 & 2m & VA\\ 
01.11.2010 & 55502.499 & 16.59 & 17.84 & 19.10 & 20.72 & 2m & VA\\ 
04.11.2010 & 55505.493 & 16.47 & -     & -     & - & Scm & FLI\\ 
05.11.2010 & 55506.484 & 16.19 & -     & -     & - & Scm & FLI\\ 
07.11.2010 & 55507.530 & 16.27 & -     & -     & - & Scm & FLI\\ 
06.01.2011 & 55568.453 & 15.81 & 17.35 & 18.90 & - & 2m & VA\\ 
08.01.2011 & 55570.298 & 15.92 & 17.42 & 19.00 & 20.82 & 2m & VA\\ 
10.01.2011 & 55571.510 & 15.93 & 17.39 & 18.91 & - & 2m & VA\\ 
11.01.2011 & 55573.304 & 16.71 & 17.96 & 19.25 & - & 2m & VA\\ 
06.02.2011 & 55599.347 & 16.51 & -     & -     & - & Sch & FLI\\ 
07.02.2011 & 55600.358 & 16.56 & -     & -     & - & Sch & FLI\\ 
04.04.2011 & 55656.278 & 16.69 & -     & -     & - & Sch & FLI\\ 
17.08.2011 & 55790.591 & 16.16 & 17.34 & 18.61 & - & 1.3m & ANDOR\\ 
18.08.2011 & 55791.607 & 15.97 & 17.23 & 18.47 & - & 1.3m & ANDOR\\ 
10.09.2011 & 55815.494 & 16.21 & 17.28 & 18.62 & - & 1.3m & ANDOR\\ 
12.09.2011 & 55816.565 & 16.09 & 17.22 & 18.39 & - & 1.3m & ANDOR\\ 
20.09.2011 & 55824.507 & 16.06 & 17.16 & -     & - & 1.3m & ANDOR\\ 
07.10.2011 & 55842.462 & 15.71 & 16.95 & 18.09 & - & 1.3m & ANDOR\\ 
13.10.2011 & 55848.460 & 15.92 & 17.11 & 18.27 & - & 1.3m & ANDOR\\ 
30.10.2011 & 55865.463 & 16.08 & 17.16 & 18.25 & - & 2m & VA\\ 
31.10.2011 & 55866.468 & 16.02 & 17.15 & 18.21 & 19.90 & 2m & VA\\ 
26.11.2011 & 55892.430 & 16.22 & 17.27 & 18.33 & - & 2m & VA\\ 
27.11.2011 & 55893.387 & 15.92 & -     & -     & - & Sch & FLI\\ 
29.11.2011 & 55895.456 & 16.01 & -     & -     & - & Sch & FLI\\ 
30.11.2011 & 55896.422 & 15.96 & 16.99 & -     & - & Sch & FLI\\ 
30.12.2011 & 55925.532 & 14.82 & -     & -     & - & Sch & FLI\\ 
01.01.2012 & 55928.327 & 15.10 & 16.56 & 17.80 & - & Sch & FLI\\ 
16.03.2012 & 56003.320 & 14.80 & 16.18 & 17.31 & - & Sch & FLI\\ 
26.03.2012 & 56013.322 & 15.87 & 16.89 & 17.95 & - & 2m & VA\\ 
28.03.2012 & 56015.326 & 15.66 & 16.78 & 17.77 & - & 2m & VA\\ 
12.04.2012 & 56030.279 & 15.87 & -     & -     & - & Sch & FLI\\ 
02.08.2012 & 56141.602 & 14.22 & 15.71 & 17.00 & - & 1.3m & ANDOR\\ 
03.08.2012 & 56142.603 & 14.27 & 15.80 & 17.15 & - & 1.3m & ANDOR\\ 
04.08.2012 & 56143.602 & 14.10 & 15.60 & 16.89 & - & 1.3m & ANDOR\\ 
14.08.2012 & 56153.608 & 14.08 & 15.63 & 16.89 & - & 1.3m & ANDOR\\ 
18.08.2012 & 56157.622 & 13.92 & 15.37 & 16.64 & - & 1.3m & ANDOR\\ 
21.08.2012 & 56160.572 & 14.27 & 15.78 & 17.10 & - & 1.3m & ANDOR\\ 
22.08.2012 & 56161.598 & 14.36 & -     & -     & - & Sch & FLI\\ 
03.09.2012 & 56173.549 & 14.20 & 15.72 & 17.04 & 18.71 & 1.3m & ANDOR\\ 
10.09.2012 & 56180.591 & 14.62 & 16.16 & 17.45 & - & 1.3m & ANDOR\\ 
11.09.2012 & 56182.500 & 14.40 & 15.94 & 17.22 & 18.87 & 1.3m & ANDOR\\ 
13.09.2012 & 56183.541 & 14.45 & 15,93 & 17.13 & 18.70 & 1.3m & ANDOR\\
23.09.2012 & 56193.542 & 14.83 & 16.31 & 17.47 & - & 1.3m & ANDOR\\
23.09.2012 & 56193.559 & 14.77 & 16.17 & -     & - & Sch & FLI\\ 
24.09.2012 & 56194.531 & 14.73 & 16.26 & 17.33 & - & Sch & FLI\\
09.10.2012 & 56209.547 & 14.73 & 16.11 & 17.23 & - & Sch & FLI\\ 
13.10.2012 & 56214.396 & 14.26 & 15.76 & 16.98 & 18.60 & 2m & VA\\
17.11.2012 & 56249.430 & 13.72 & 15.13 & 16.30 & 17.97 & Sch & FLI\\
14.12.2012 & 56275.562 & 14.11 & 15.56 & 16.83 & 18.49 & 2m & VA\\ 
14.12.2012 & 56276.414 & 14.49 & 15.95 & 17.17 & 18.82 & 2m & VA\\
19.01.2013 & 56312.357 & 15.00 & 16.48 & 17.66 & - & 2m & VA\\ 
04.02.2013 & 56328.438 & 14.72 & 16.09 & 17.18 & - & Sch & FLI\\ 
05.02.2013 & 56329.422 & 14.79 & 16.14 & 17.25 & - & Sch & FLI\\
04.03.2013 & 56356.373 & 14.88 & 16.17 & -     & - & 60cm & FLI\\
17.03.2013 & 56369.341 & 14.34 & 15.81 & 17.21 & - & 2m & VA\\ 
19.03.2013 & 56371.332 & 14.15 & -     & -     & - & 2m & VA\\ 
10.04.2013 & 56393.277 & 14.45 & 16.05 & 17.30 & - & Sch & FLI\\ 
03.08.2013 & 56507.572 & 13.88 & -     & -     & - & 2m & VA\\
04.08.2013 & 56508.573 & 13.57 & 14.94 & 16.20 & - & 2m & VA\\
05.08.2013 & 56509.575 & 13.54 & 14.98 & -     & - & Sch & FLI\\ 
06.08.2013 & 56510.573 & 13.54 & 14.94 & 16.13 & - & Sch & FLI\\ 
07.08.2013 & 56511.574 & 13.65 & -     & -     & - & Sch & FLI\\ 
08.08.2012 & 56512.571 & 13.83 & 15.25 & 16.41 & - & Sch & FLI\\
05.09.2013 & 56540.564 & 13.75 & 15.20 & 16.44 & - & Sch & FLI\\
06.09.2013 & 56541.546 & 13.61 & 15.04 & 16.23 & - & Sch & FLI\\ 
07.09.2013 & 56542.537 & 13.78 & 15.24 & 16.45 & 18.06 & Sch & FLI\\ 
09.09.2013 & 56544.528 & 14.01 & 15.42 & 16.73 & 18.40 & 2m & VA\\ 
17.09.2013 & 56553.525 & 13.39 & 14.76 & 15.95 & 17.55 & 1.3m & ANDOR\\
13.10.2013 & 56578.592 & 13.46 & 14.82 & -     & - & 60cm & FLI\\ 
08.11.2013 & 56604.545 & 13.35 & 14.70 & -     & - & 60cm & FLI\\ 
09.12.2013 & 56636.332 & 13.31 & 14.70 & 15.82 & 17.42 & 2m & VA\\ 
28.12.2013 & 56655.432 & 13.21 & 14.54 & 15.66 & 17.30 & Sch & FLI\\ 
30.12.2013 & 56656.517 & 13.29 & 14.65 & 15.84 & 17.43 & Sch & FLI\\
30.12.2013 & 56657.279 & 13.45 & 14.81 & -     & - & Sch & FLI\\ 
23.01.2014 & 56681.461 & 14.10 & 15.63 & 16.89 & - & Sch & FLI\\ 
05.02.2014 & 56694.412 & 14.36 & 15.67 & 17.17 & 19.12 & 2m & VA\\ 
21.03.2014 & 56738.291 & 13.98 & 15.54 & 16.79 & 18.50 & Sch & FLI\\ 
23.03.2014 & 56740.302 & 13.97 & 15.46 & 16.72 & - & Sch & FLI\\ 
30.08.2014 & 56899.529 & 13.26 & 14.60 & 15.77 & 17.36 & 1.3m & ANDOR\\ 
26.11.2014 & 56988.337 & 13.62 & 15.07 & 16.29 & 18.09 & Sch & FLI\\ 
14.12.2014 & 57005.532 & 13.67 & 15.05 & 16.20 & - & Sch & FLI\\ 
15.12.2014 & 57006.576 & 13.44 & 14.77 & 15.88 & - & Sch & FLI\\
25.12.2014 & 57016.556 & 13.91 & 15.36 & 16.62 & 18.25 & 2m & VA\\ 
25.12.2014 & 57017.346 & 13.88 & 15.30 & 16.55 & 18.16 & 2m & VA\\
18.02.2015 & 57072.365 & 13.34 & 14.60 & 15.62 & 17.14 & Sch & FLI\\ 
20.02.2015 & 57074.361 & 13.66 & 14.94 & 16.01 & 17.59 & Sch & FLI\\ 
23.04.2015 & 57136.255 & 13.22 & 14.48 & 15.57 & 17.09 & Sch & FLI\\
\end{longtable}

\end{document}